\documentstyle[epsf, rotate]{article}
\begin{document}
\begin{center}
{\Large
{\bf Spatial distribution of accreting isolated neutron stars in the Galaxy}
}\\

\vskip 0.3cm

S.B.Popov and M.E.Prokhorov\\

Sternberg Astronomical Institute, Moscow State University\\

\end{center}

\begin{abstract}

    We present here the computer model of the
distribution of the luminosity,
produced by old isolated neutron stars (OINSs)
accreting from the interstellar medium (ISM).
We show, that for different mean velocities of OINSs
the distribution of the luminosity
has a torus-like structure, with the maximum at $\approx 5 kpc$.

\end{abstract}

\section{Introduction}

	In the last several years, the spatial distribution
of old isolated neutron stars (OINSs) became of great interest
(see, for example, Treves and Colpi (1991)).
Several sources of this size were observed by ROSAT.
Different regimes of interaction of the
interstellar medium (ISM) and OINSs can appear: Ejector,
Propeller (with possible transient source), Accretor,
Georotator and supercritical regimes
(see, Popov (1994) and Lipunov and Popov (1995)).
Here we are interested only in accreting OINSs.

We use direct calculations of trajectories
in the Galaxy potential, taken in the form (Paczynski  (1990) ):

$$
    \Phi_i (R,Z)=GM_i/\left(R^2+[a_i+(Z^2+b_i^2)^{1/2}]^2\right)^{1/2}
$$
.

	In the articles of Postnov and Prokhorov (1993, 1994) it was shown,
that OINSs in the Galaxy form a torus-like structure. If one looks at their
distribution and at the distribution of the ISM (see, for example,
Bochkarev (1993) ), it is clearly seen, that the maximums of two distributions
roughly coincides. It means, that most part of OINSs is
situated in dence regions of ISM.
So, the luminosity there must be higher.
Here we represent  computer simulations of this
 situation.

\begin{figure}
\epsfxsize=8cm
\centerline{\rotate[r]{\epsfbox{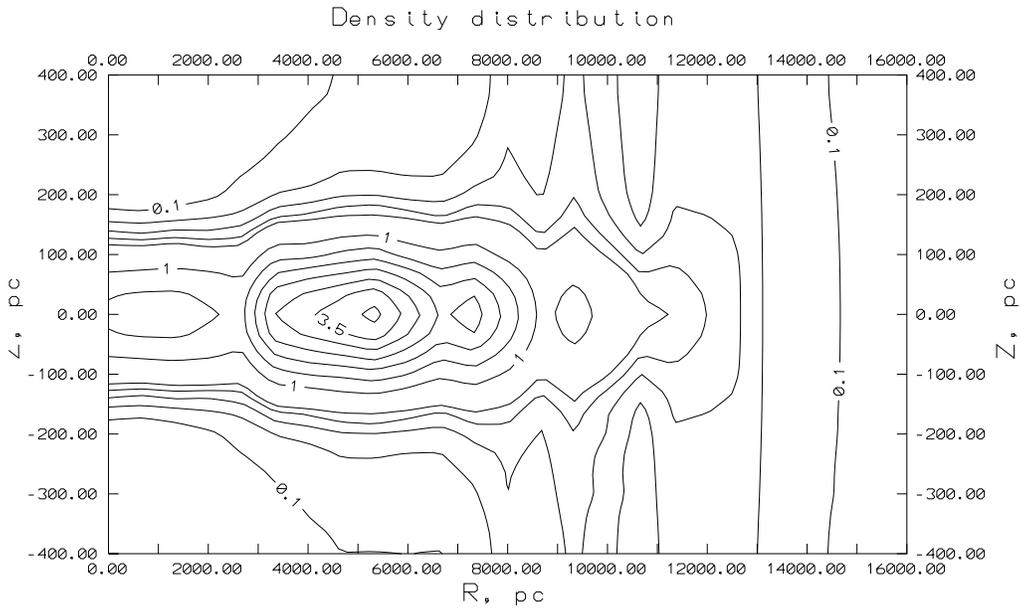}}}
\caption{ The density distribution in R-Z plane }
\end{figure}

\begin{figure}
\epsfxsize=7cm
\centerline{\rotate[r]{\epsfbox{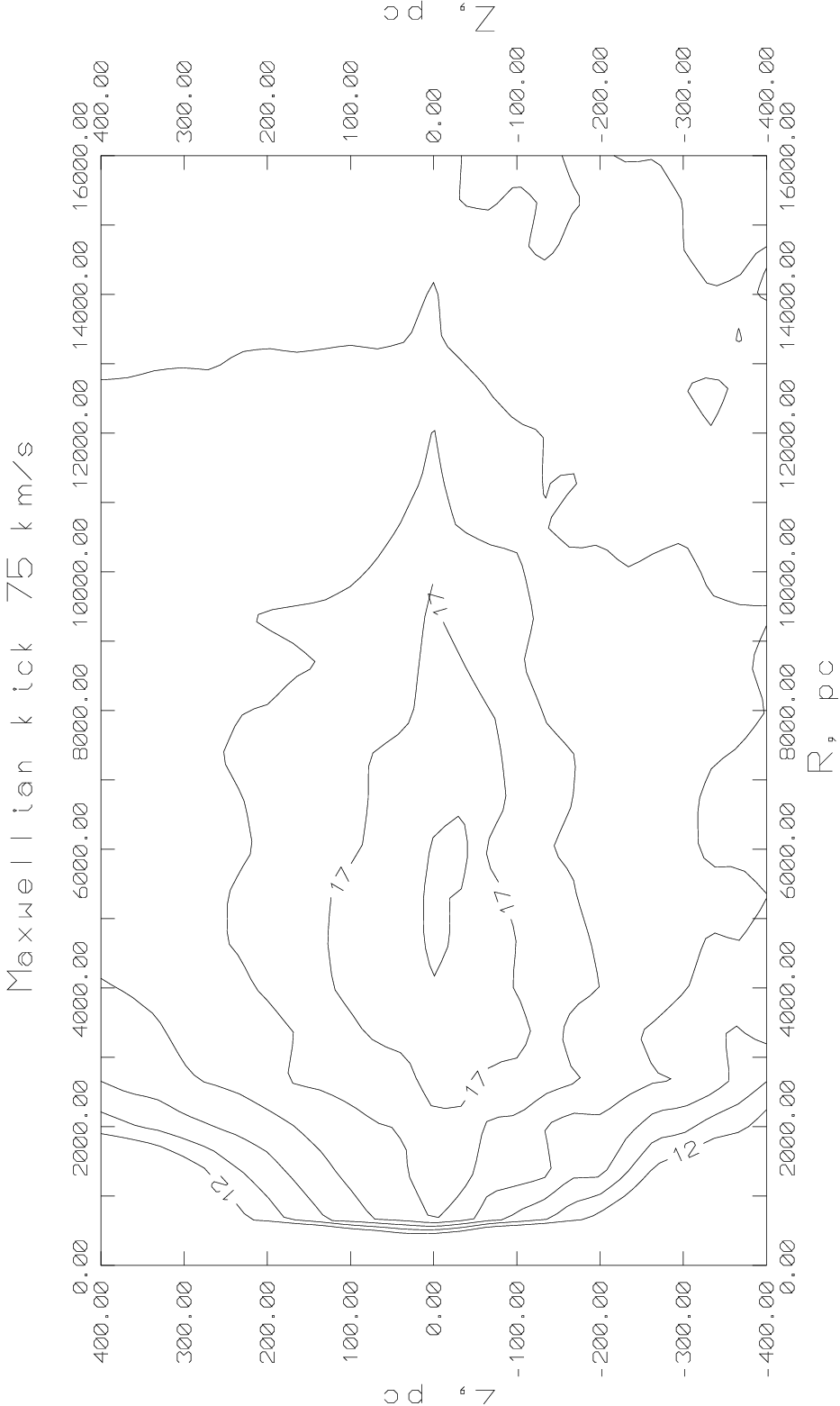}}}
\caption{ The luminosity distribution in R-Z plane for
Maxwellian kick velocity (75 km/s)}
\end{figure}

\begin{figure}
\epsfxsize=7cm
\centerline{\rotate[r]{\epsfbox{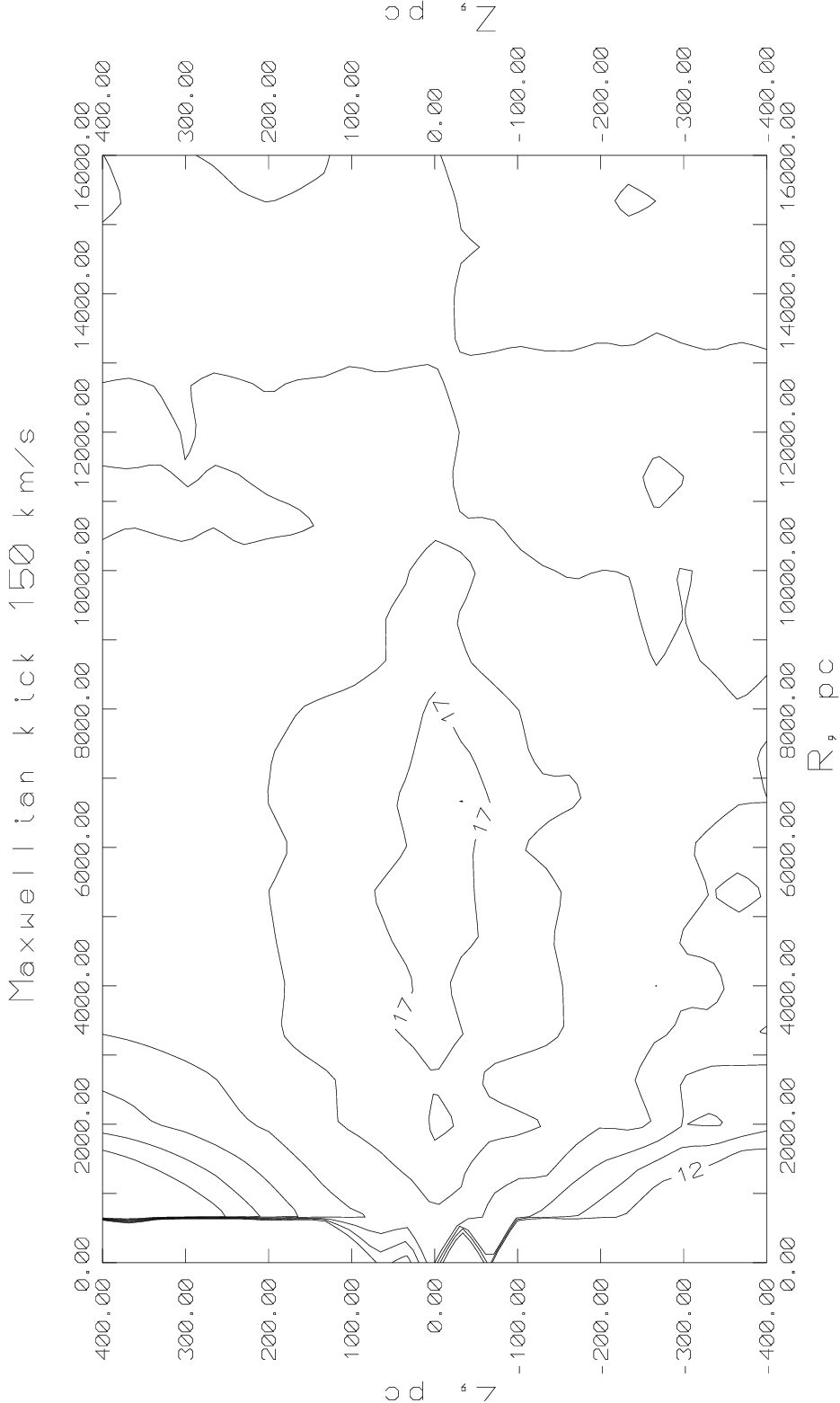}}}
\caption{ The luminosity distribution in R-Z plane for
Maxwellian kick velocity (150 km/s)}
\end{figure}

\section{Model}

    We made calculations on the grid with the cell size 100 pc in R-direction
and 10 pc in Z-direction (centered at R=50 pc, Z=5 pc and so on).
Stars were born in the Galactic plane.
The system of differential equations was solved numerically.

	In our model we assumed, that the birthrate of NSs is
proportional to the square of local density. Local density was calculated
using data and formulaes from Bochkarev (1993) and Zane et al. (1995).

$$
   n(R, Z)= n_{HI}+2\cdot n_{H_2}
$$

$$
   n_{H_2}=n_0\cdot exp \left[ \frac{ -Z^2}{2\cdot (70 pc)^2} \right]
$$

   If $2 kpc \le R \le 3.4 kpc $, then

$$
   n_{HI}=n_0\cdot exp \left[ \frac{-Z^2}{2\cdot (140 pc \cdot R/ 2 kpc)^2} \right],
$$

	For $R \le 2 kpc \, \,  n(R,Z) $ was  assumed to be constant:

$$
   n(R<2 kpc, Z)=n(R=2 kpc, Z)
$$
    Of course, it is not accurate, so for the very central part
of the Galaxy our results are only a rough estimation.

    If $3.4 kpc \le R \le 8.5 kpc$, then

$$
   n_{HI}=0.345\cdot exp \left[ \frac{-Z^2}{2\cdot (212 pc)^2} \right] +
          0.107\cdot exp \left[ \frac{-Z^2}{2\cdot (530 pc)^2} \right] +
          0.064\cdot exp \left[ \frac{-Z}{403 pc} \right]
$$

    If $ 8.5 \le R \le 16 kpc$, then

$$
   n_{HI}=n_{\infty}\cdot exp \left[ \frac{-Z^2}{2\cdot (530 pc \cdot R / 8.5 kpc)^2}
\right]
$$

    The density distribution is shown in the figure 1.

\begin{figure}
\epsfxsize=6cm
\centerline{\rotate[r]{\epsfbox{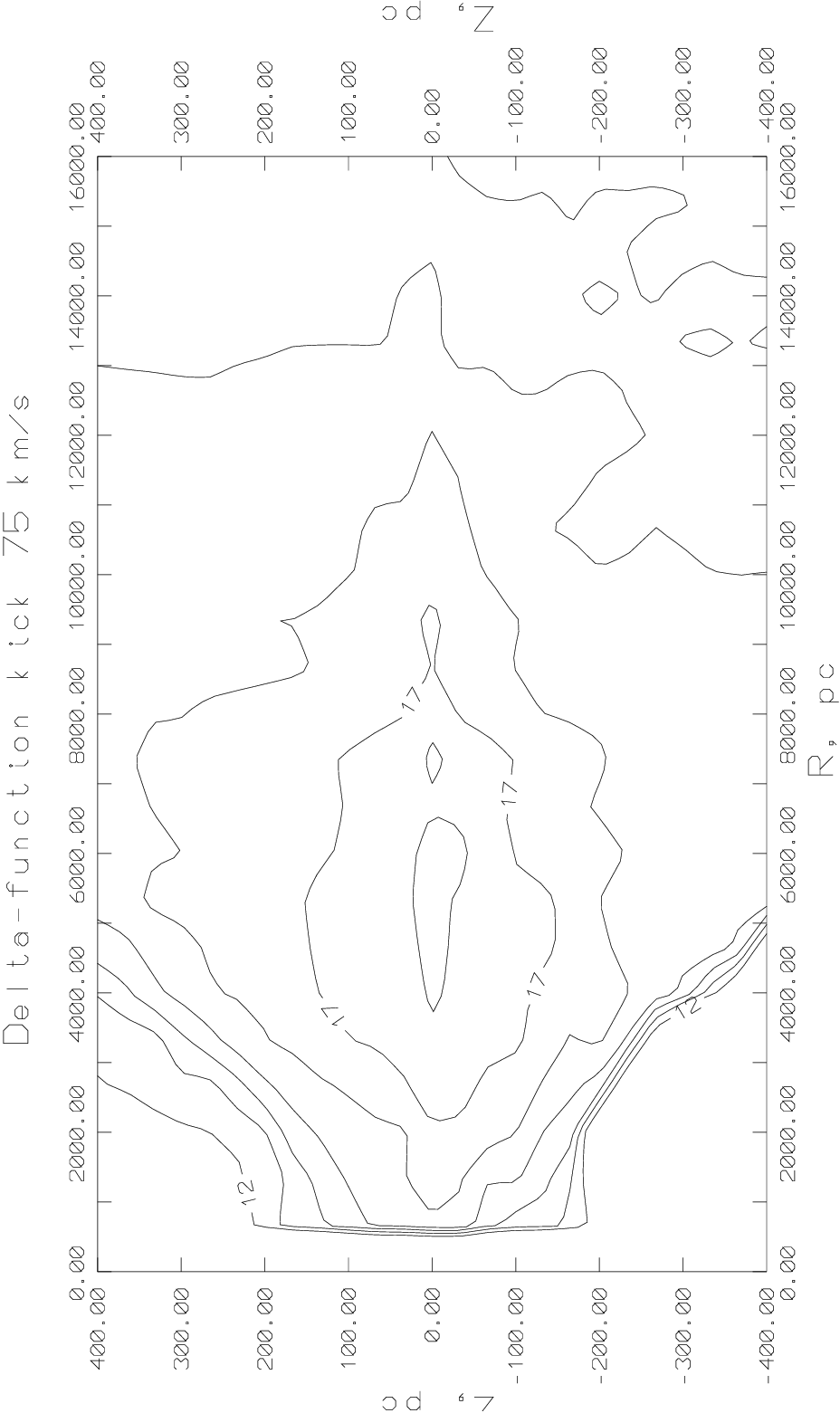}}}
\caption{The luminosity distribution in R-Z plane for
$\delta$-function kick velocity (75 km/s)}
\end{figure}

\begin{figure}
\epsfxsize=7cm
\centerline{\rotate[r]{\epsfbox{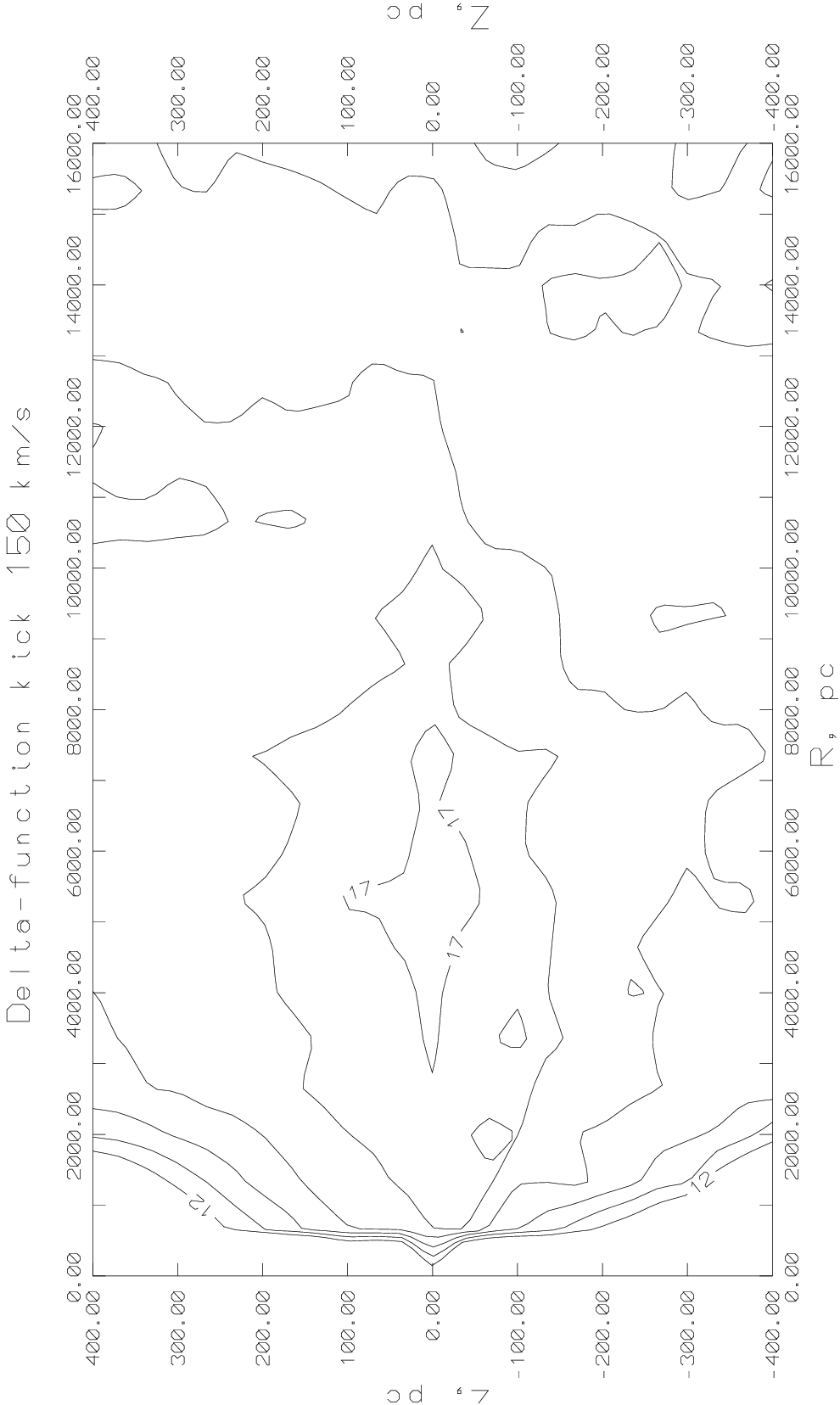}}}
\caption{The luminosity distribution in R-Z plane for
$\delta$-function kick velocity (150 km/s)}
\end{figure}

\begin{figure}
\epsfxsize=7cm
\centerline{\rotate[r]{\epsfbox{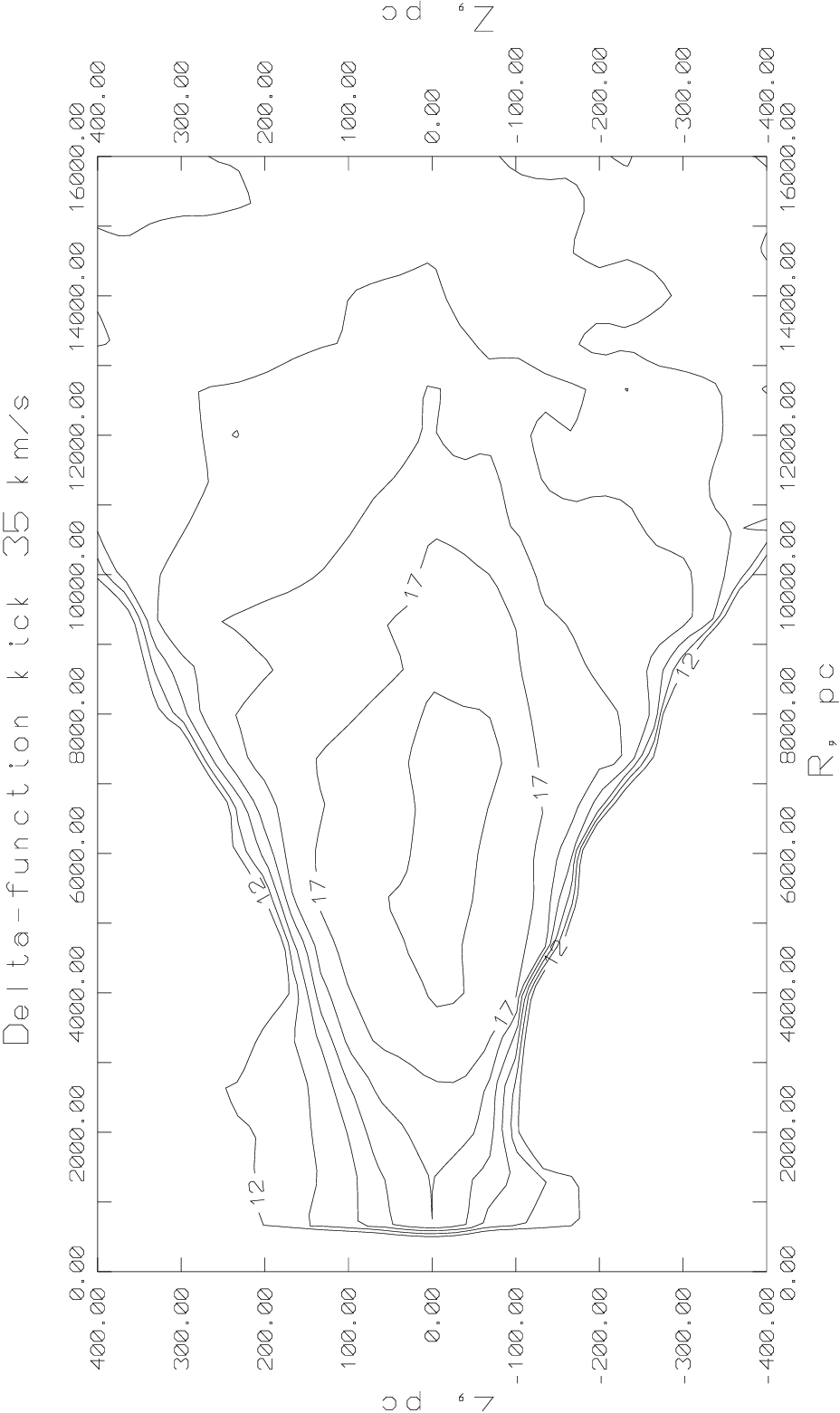}}}
\caption{The luminosity distribution in R-Z plane for
$\delta$-function kick velocity (35 km/s)}
\end{figure}

\begin{figure}
\epsfxsize=7cm
\centerline{\rotate[r]{\epsfbox{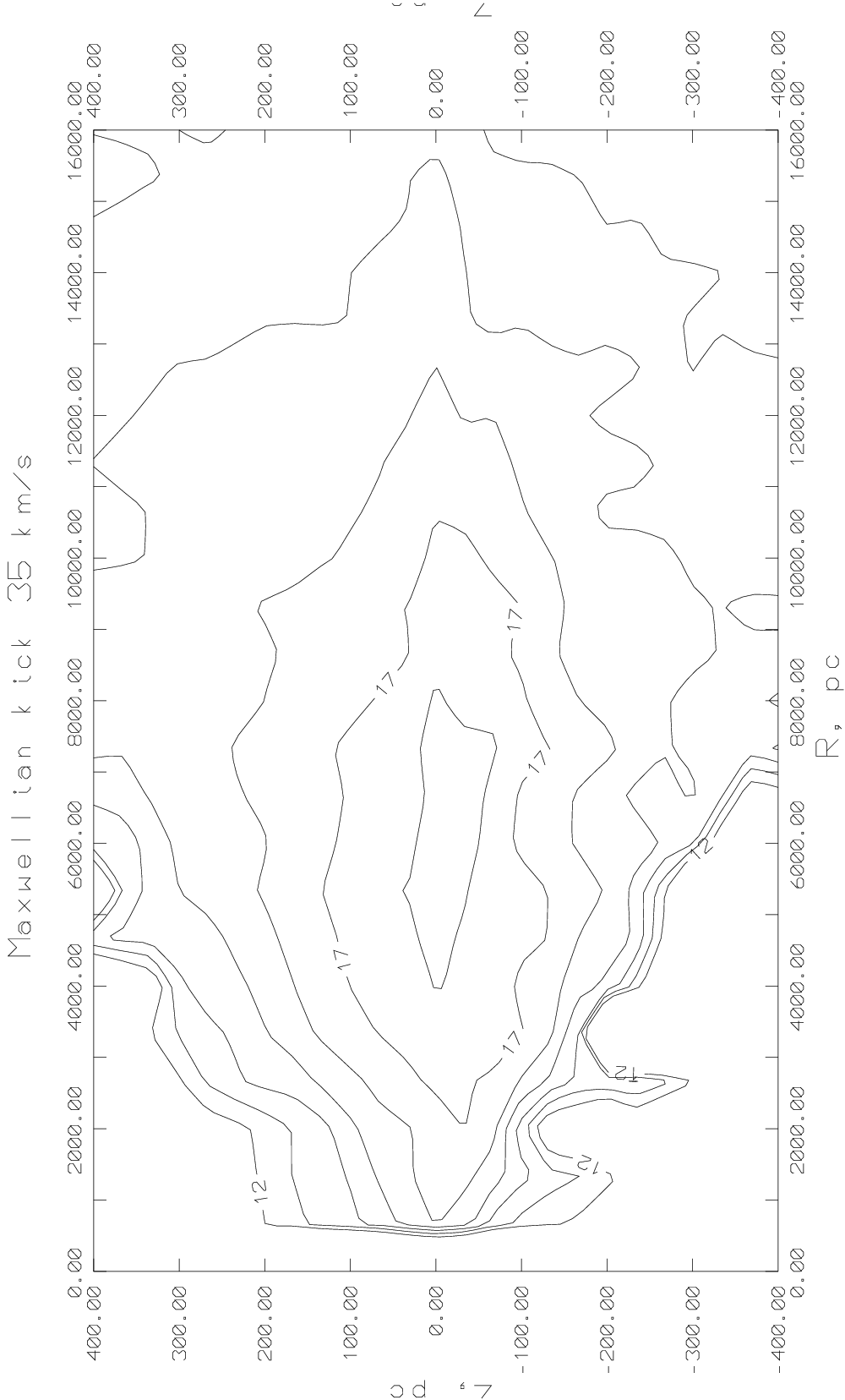}}}
\caption{The luminosity distribution in R-Z plane for
maxwellian kick velocity (35 km/s)}
\end{figure}

\begin{figure}
\epsfxsize=7cm
\centerline{\rotate[r]{\epsfbox{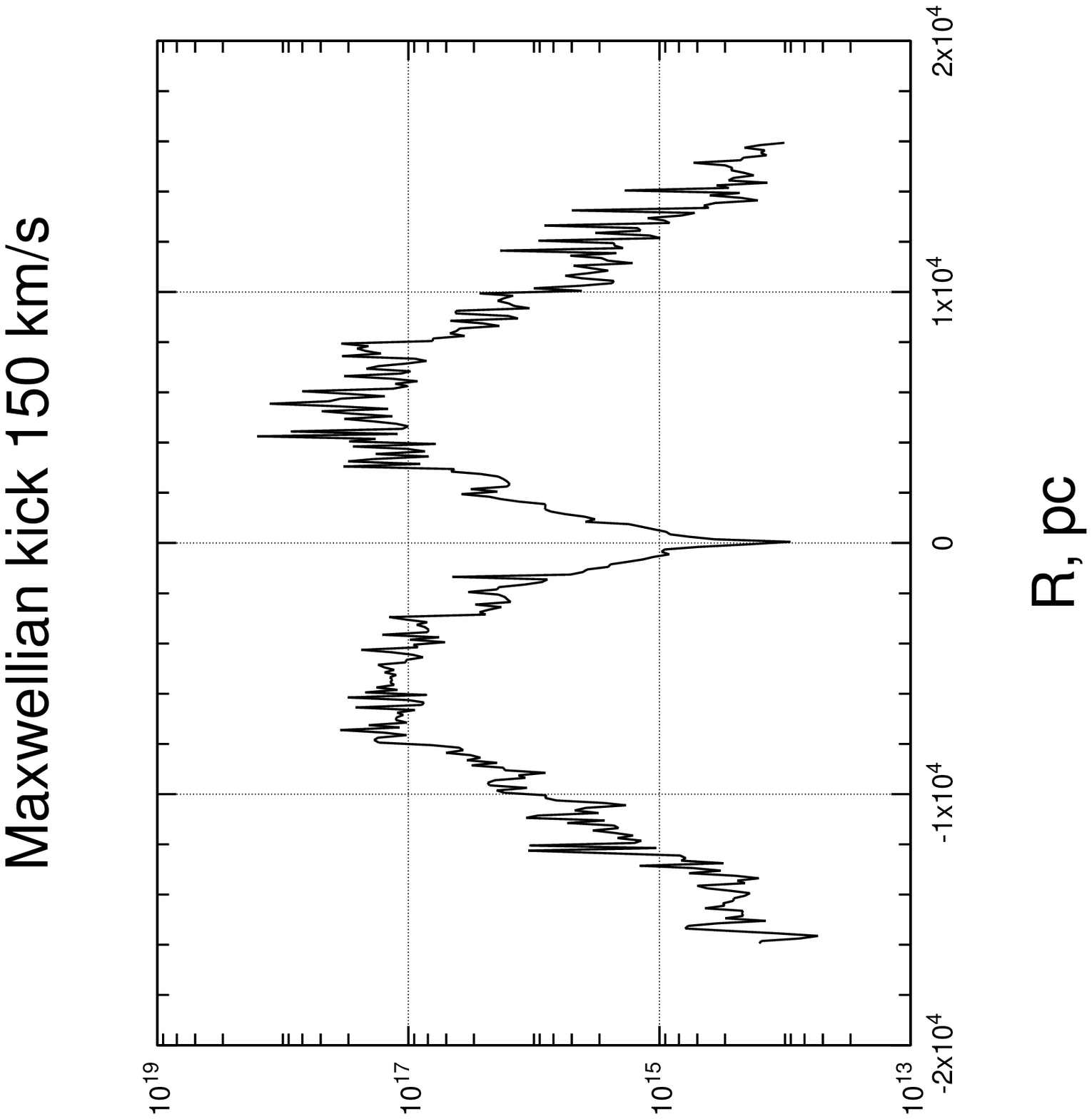}}}
\caption{Slice at Z=+5 pc for maxwellian kick velocity (150 km/s) }
\end{figure}

Kick velocity was taken both: in the Maxwellian form
with the maximum velocity 150 km/s, 75 km/s and 35 km/s
and as a $\delta$-function with V=150 km/s, 75 km/s  and 35 km/s
(see discussion in Lipunov et al. (1996) ).

    Sound velocity was taken to be 10 km/s.
Luminosity was calculated using Bondi formula:

$$
    L=\left(\frac{GM_{NS}}{R_{NS}}\right) 2 \pi \left(
    \frac{(GM_{NS})^2 n(R,Z)}{(V_s^2+V^2)^{3/2}}\right)
$$
.

\section{Results}

    On the figures 2-7 we represent the results for two velocity
distributions. On the figure 8 the slice at Z=+5 pc
for the maxwellian kick ($V_{max}$= 150 km/s) is shown.

    As it is clearly seen from the figures, the distribution of
the luminosity density (shown in arbitrary units) in R-Z plane
forms a torus-like structure with the maximum
at approximatelly 5 kpc.

\section{Discussion and concluding remarks}

    The torus-like structure of that distribution
is an interesting and important feature
of the Galactic potential. Local maximums in the ISM distribution
are smoothed (compare figures 1-7). As one can suppose, for low
velocities we get greater luminosity. Stars with
the Maxwellian distribution can penetrate deeper into the inner
regions than stars with $\delta$-function velocity distribution
(especially it is clear for high Z - 200-400 pc for low velocity
distributions) because we for maxwellian kick we have both:
more low velocity and more high velocity stars.

    On fig.9 we show dependence of the total luminosity of the galaxy
(in arbitrary units) from the kick velocity for two types of distributions.
We mark very interesting feature: intersection of the curves at
$\approx 125$ km/s.

\begin{figure}
\epsfxsize=7cm
\centerline{\rotate[r]{\epsfbox{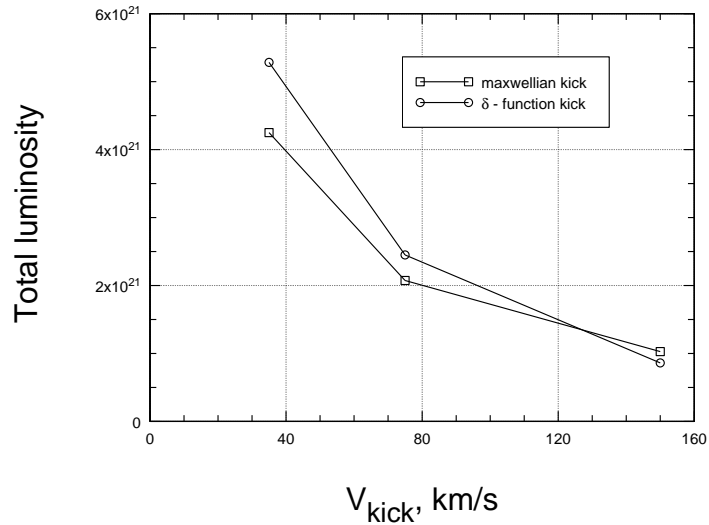}}}
\caption{Total velocity (in arbitrary units) vs. kick velocity }
\end{figure}

    As me made very general assumptions, we argue, that such a distribution
is not unique for our Galaxy, and all spiral galaxies must have such
a distribution of the luminosity density, associated with accreting OINSs.

\section{Aknowledgements}

    The work was supported by the RFFI (95-02-6053) and
the INTAS (93-3364) grants.
The work of S.P. was also supported by the ISSEP.



\begin{thebibliography}{}

\bibitem{} N.G. Bochkarev, "Basics of the ISM physics",
1992, Moscow, Moscow State Univ. Press

\bibitem{} V.M. Lipunov and S.B. Popov, AZh, 71, 711, 1995

\bibitem{} V.M. Lipunov, K.A. Postnov and M.E. Prokhorov, A\&A, 310, 489, 1996

\bibitem{} B. Paczynski, ApJ 348, 485, 1990

\bibitem{} S.B. Popov, Astr. Circ., N1556, 1, 1994

\bibitem{} M.E. Prokhorov and K.A. Postnov, A\&A, 286, 437, 1994

\bibitem{} M.E. Prokhorov and K.A. Postnov,  Astr. Astroph. Trans., 4, 81, 1993

\bibitem{} A. Treves and M. Colpi, A\&A, 241, 107, 1991

\bibitem{} S. Zane, R. Turolla, L. Zampieri, M. Colpi and A. Treves,
           ApJ, 1995, 451, 739


\end{thebibliography}
\end{document}